\documentclass[12pt, aps, prb, onecolumn, nofootinbib, superscriptaddress, showkeys, floatfix, amsmath, amsfonts, amssymb, amsthm]{revtex4-2}

\usepackage[utf8]{inputenc}
\usepackage[T1]{fontenc}
\usepackage[main=english]{babel}

\usepackage{setspace}\setdisplayskipstretch{}
\usepackage{mleftright}

\usepackage{bm} 
\usepackage{mathrsfs} 
\usepackage{bbm} 
\usepackage{upgreek} 

\usepackage{hyperref}
\hypersetup{colorlinks=true, linkcolor=blue, citecolor=blue, urlcolor=blue, filecolor=red, pdfstartview=}

\makeatletter
\newcommand*{\defeq}{\mathrel{\rlap{%
\raisebox{0.3ex}{$\m@th\cdot$}}%
\raisebox{-0.3ex}{$\m@th\cdot$}}=}
\makeatother

\newcommand*{\iu}{\mathrm{i}} 
\newcommand*{\Elr}{\mathrm{e}} 

\newcommand*{\abs}[1]{\mleft\lvert {#1} \mright\rvert} 
\newcommand*{\dd}[2][]{\mathop{}\!\mathrm{d}^{#1} {#2}} 
\newcommand*{\var}[2][]{\mathop{}\!\delta_{#1} {#2}} 

\newcommand*{\vdot}{\bm{\cdot}} 
\newcommand*{\vb}[1]{\bm{#1}} 

\newcommand*{\ev}[1]{\mleft\langle {#1} \mright\rangle} 

\newcommand*{\Haml}{\mathcal{H}} 

\newcommand*{\qev}[1]{\mleft\langle \hspace{-3pt}\middle\langle {#1} \middle\rangle \hspace{-3pt}\mright\rangle}

\begin{document}

\title{Physical limitations of the Hohenberg-Mermin-Wagner theorem}

\author{Grgur Palle}
\affiliation{Department of Physics, Faculty of Science, University of Zagreb}
\affiliation{Institute for Theoretical Condensed Matter Physics, Karlsruhe Institute of Technology}
\author{D K Sunko}
\affiliation{Department of Physics, Faculty of Science, University of Zagreb}

\begin{abstract}
{The Hohenberg-Mermin-Wagner (HMW) theorem states that infrared (IR)
fluctuations prevent long-range order which breaks continuous symmetries in two
dimensions (2D), at finite temperatures.} We note that the theorem becomes
physically effective for superconductivity (SC) only for astronomical sample
sizes, so it does not prevent 2D SC in practice. We systematically explore the
sensitivity of the magnetic and SC versions of the theorem to finite-size and
disorder effects. For magnetism, finite-size effects, disorder, and
perpendicular coupling can all restore the order parameter at a non-negligible
value of $T_c$ equally well, making the physical reason for finite $T_c$
sample-dependent. For SC, an alternative version of the HMW theorem is
presented, in which the temperature cutoff is set by Cooper pairing, in place of
the Fermi energy in the standard version. It still allows 2D SC at $2$--$3$
times the room temperature when the interaction scale is large and Cooper pairs
are small, the case with high-$T_c$ SC in the cuprates. Thus IR fluctuations do
not prevent 2D SC at room temperatures in samples of any reasonable size, by any
known version of the HMW argument. A possible approach to derive
mechanism-dependent upper bounds for SC $T_c$ is pointed out.
\end{abstract}


\maketitle

\section{Introduction}

The Hohenberg-Mermin-Wagner (HMW) theorem~\cite{Hohenberg,Mermin} is probably
the best-known mathematically exact result in the theory of phase transitions.
It forbids ordered phases which break continuous symmetries in less than three
dimensions, by showing that they are destabilized by infrared (IR) fluctuations.
The formal prerequisite for the argument is due to
Bogoliubov~\cite{Bogoliubov62}: because the {static} susceptibility of a
many-body system can be written as a scalar product in a certain operator vector
space, the Cauchy-Schwartz inequality in that space yields rigorous inequalities
for the {static} response functions of arbitrary operators. With well-chosen probes
and a summation over momenta~\cite{Hohenberg,Halperin}, the theorem follows.

Despite its wide dissemination in textbooks and the research
literature~\cite{Gelfert-Nolting,Kuzemsky,Sadovnikov,Gelfert,Halperin}, we have
been able to find only one comment, in unpublished lecture notes by
Leggett~\cite{Leggett-lecture}, on the actual numbers appearing when the
theoretical bound is evaluated: a 2D sample would have to be ``the size of the
Moon's orbit'' for its superconductivity (SC) to be suppressed below the
temperatures at which it is observed in three dimensions. The reason is easy to
spot. In the formalism, the energy cost of the IR fluctuations suppressing $T_c$
is not set by the SC mechanism, as one might imagine, but by the much larger
Fermi energy. Because the HMW suppression is only logarithmic in the size of the
sample in 2D, it cannot preclude 2D SC in reasonably-sized samples even at twice
the room temperature.

Two-dimensional SC is of considerable practical interest since the discovery of
\linebreak high-$T_c$ SC in cuprates~\cite{Bednorz86}. These are strongly
anisotropic materials with ionic layers between 2D metallic planes. Later, 2D
heterostructures were fabricated in which a \emph{single} plane was
superconducting without any suppression of
$T_c$~\cite{cuprate-monolayer1,cuprate-monolayer2}. Similarly, SC thin films of
FeSe were found to superconduct at
$\sim$~$70$~K~\cite{FeSe-monolayer1,FeSe-monolayer2}. In these cases, one could
still harbor the suspicion that the insulator adjacent to the active layer
somehow helped to stabilize SC. Most recently, such reservations became
implausible by the observation of undiminished SC at $\sim$~$100$~K in
exfoliated BSCCO~\cite{cuprate-monolayer4}.

Concerning the magnetic version of the HMW theorem, a number of authors have
discussed difficulties in observing suppression of magnetism in 2D in concrete
settings, both experimentally~\cite{Jongh,Pomerantz} and
numerically~\cite{RFIO1}. One way to put these issues on a physical footing is
to use disorder to recover the order parameter in 2D in a controllable
manner~\cite{RFIO1,RFIO2,RFIO3,RFIO4,RFIO5}. Such disorder-induced order is
expected because the HMW theorem assumes an exact continuous symmetry of the
Hamiltonian, which is manifestly broken by a random field. The surprise in
practice was the fragility of the HMW result: even a small amount of disorder
recovered the ordered state. However, the exact formal argument~\cite{RFIO5}
for such random-field-induced order (RFIO) was specialized to the classical
random-field XY model, leaving the general case to be inferred.

In this work, we first analyze the HMW theorem for the XXZ model with disorder,
establishing a generic rationale for the above-mentioned observations for
magnetism. Our investigation of possible ways to gap the IR fluctuations,
required by the magnetic HMW mechanism, does not prejudice other mechanisms. In
particular, the so-called Imry-Ma argument~\cite{Imry75} may (or may
not~\cite{RFIO5,Changlani2016,Berzin2017}) be relevant in a given physical
situation. Next, we show why the original HMW formulation for SC is robustly
independent of model details, as long as the interaction is local, or almost
local, in real space. This observation encompasses all variations of the HMW
theorem proven separately over the
years~\cite{Sadovnikov,Gelfert,supra1,supra2,supra3,supra4,supra5}. After that,
we present a new variation of the HMW argument which probes the interaction
scale, instead of the Fermi energy. It is valid in any dimension, and
independent of the size of the sample. Rough evaluation in 2D gives upper limits
of $T_c$ on the high end, comparable to the finite-size HMW bound, when Cooper
pairs are small. The approach is still based on the amplitude of the order
parameter, which is qualitatively different from the reduction $k_B T_c < E_F /
8$ of the infinite-size HMW bound, obtained by considering phase
stiffness~\cite{alt-Tc-bounds}. Finally, we point out a possible way forward, to
derive mechanism- and material-dependent upper bounds for the SC $T_c$ by
refining the present reasoning.

\section{Finite-size and disorder effects}

\subsection{Finite-size effects in the XXZ model}

{Let us consider a system of localized spins described by the XXZ
model:
\begin{equation}
\Haml = - \sum_{ij} J_{ij} \mleft(S_{i}^{x} S_{j}^{x} + S_{i}^{y}
S_{j}^{y}\mright) - \sum_{ij} J_{ij}^{z} S_{i}^{z} S_{j}^{z} - B \, V M_x\,,
\label{eq:XXZ-Haml}
\end{equation}
where $B$ is the external symmetry-breaking field and $M_x = V^{-1} \sum_{i}
\Elr^{- \iu \vb{K} \vdot \vb{R}_i} S_{i}^{x}$ the staggered magnetization. The
Bogoliubov inequality
\begin{equation}
\abs{\ev{[A, Q]}}^2 \leq \frac{1}{2} \beta \, \ev{\{A, A^{\dag}\}} \,
\ev{\mleft[[Q, \Haml], Q^{\dag}\mright]}\,,
\label{Bglj-ineq}
\end{equation}
where braces are anticommutators, $\beta$ is the inverse temperature, and $Q$
and $A$ are arbitrary operators. To probe IR fluctuations of some broken
symmetry, $Q$ is chosen to be a modulation of its generator and $A$ is tuned so
that $\ev{[A, Q]}$ gives the corresponding order parameter. For the XXZ model,
the appropriate choices are $Q_{\vb{k}} = \sum_{i} \Elr^{- \iu \vb{k}
\vdot \vb{R}_i} S_{i}^{z}$ and $A_{\vb{k}, \vb{K}} = V^{-1} \sum_{i} \Elr^{\iu
(\vb{k} - \vb{K}) \vdot \vb{R}_i} S_{i}^{y}$ which yield $ [A_{\vb{k},
\vb{K}}, Q_{\vb{k}}] = \iu M_{x}$.

To obtain from~(\ref{Bglj-ineq}) a useful bound on $\ev{M_x}_B$, one has to
divide~(\ref{Bglj-ineq}) by $\ev{\mleft[[Q, \Haml], Q^{\dag} \mright]}$ and sum
it over momenta $\vb{k}$ of the first Brillouin zone. That way the orthogonality
of the $\Elr^{\iu (\vb{k} - \vb{K}) \vdot \vb{R}_i}$ factors from $A_{\vb{k},
\vb{K}}$ ensures the \emph{intensive} bound $\sum_{\vb{k} \in
\mathrm{BZ}} \langle \{A_{\vb{k}, \vb{K}}, A^{\dag}_{\vb{k}, \vb{K}}\}
\rangle_B \leq 2 n^2 S^2$, where $n = \mathcal{N} / V$ is the concentration
and $\mathcal{N}$ the number of unit cells. That $Q_{\vb{k}}$ is a generator of
symmetry implies that $\big[[Q_{\vb{k}}, \Haml], Q^{\dag}_{\vb{k}}\big]
\propto \vb{k}^2$, and indeed a rigorous bound $V^{-1} \langle [[Q_{\vb{k}},
\Haml], Q^{\dag}_{\vb{k}}] \rangle_B \leq n J (\vb{k}^2 /  k_{\mathrm{BZ}}^2)
+ \abs{B \ev{M_x}_B}$ can be derived in which $k_{\mathrm{BZ}}$ is the effective
radius of the first Brillouin zone and $J = \frac{S (S + 1)}{\mathcal{N}}
\sum_{ij} k_{\mathrm{BZ}}^2 (\vb{R}_i - \vb{R}_j)^2
\abs{J_{ij}}$ the effective spin-spin coupling constant. Thus we obtain the
HMW inequality~\cite{Mermin}
\begin{equation}
\abs{\ev{M_x}_B}^2 \leq \frac{\beta n S^2}{\displaystyle
\frac{1}{\mathcal{N}} \sum_{\vb{k} \in \mathrm{BZ}} \frac{1}{n J (\vb{k}^2 / 
k_{\mathrm{BZ}}^2) + \abs{B \ev{M_x}_B}}}\,. \label{4:Malmost}
\end{equation}
By taking the thermodynamic limit $V \to \infty$, the sum in the denominator
diverges in the limit $B \to 0$ in 1D and 2D, forbidding a finite value of the
magnetization $\ev{M_x}_0$ that is within the XY plane of symmetry.

Although the use of periodic boundary conditions apparently limits the above
argument to very large systems, finite, and even fractal~\cite{fractal-mag},
lattices can be treated with minimal technical modification. One simply
introduces a basis $\varphi_{\vb{k}}(\vb{R}_i)$ in $A_{\vb{k}, \vb{K}}$ and
$C_{\vb{k}}$ which is not a plane wave, but tuned to the
lattice~\cite{fractal-mag}. The key formal properties of
$\varphi_{\vb{k}}(\vb{R}_i)$ for the above argument are that
$\varphi_{\vb{k}}(\vb{R}_i) \to 1$ as $\vb{k} \to \vb{0}$, and that the density
of the discrete wave-vectors $\vb{k}$ near $\vb{0}$ grows sufficiently fast in
the thermodynamic limit. Thus Eq.~(\ref{4:Malmost}) still holds, with an
appropriate understanding of $\sum_{\vb{k} \in \mathrm{BZ}}$ and
$k_{\mathrm{BZ}}$, and the only physical effect of finiteness on the HMW
arguments is the appearance of an infrared cutoff, namely $k_{\mathrm{min}} =
k_{\mathrm{BZ}}/\sqrt{\mathcal{N}}$ in 2D. The sum in the denominator can then
be replaced with an integral for finite 2D samples as well,}
\begin{equation}
\frac{1}{\mathcal{N}} \sum_{\vb{k}} \to \frac{1}{\mathcal{N}} \Bigg|_{\vb{k}
= \vb{0}} + \frac{2}{k_{\mathrm{BZ}}^2}
\int_{k_{\mathrm{min}}}^{k_{\mathrm{BZ}}} \dd{k} \, k\,,
\end{equation}
turning Eq.~(\ref{4:Malmost}) into an inequality with $\abs{\ev{M_x}_B}$ on both
sides. This inequality is equivalent to $\abs{\ev{M_x}_B} \leq
M_{\mathrm{max}}$, where $M_{\mathrm{max}}$ is determined by a transcendental
equation which can be solved in the limit $B \to 0$, giving:
\begin{equation}
M_{\mathrm{max}}^2 \approx \frac{2 \beta J}{\ln(\mathcal{N}/2) + \ln(\beta J)}
\cdot M_{\mathrm{sat}}^2\,.
\end{equation}
In 1D, the corresponding $B \to 0$ solution is given by
\begin{equation}
M_{\mathrm{max}}^3 \approx \frac{4 \beta J}{\pi^2 \sqrt{\mathcal{N}}} \cdot
M_{\mathrm{sat}}^3\,,
\end{equation}
where $M_{\mathrm{sat}} = n S$ is the saturation magnetization. In the
derivation of the above, let us only note that one needs to use a finite, but
physically infinitesimal, value of $B$ to probe symmetry breaking. Technically,
we have used $B = \sqrt{\mathcal{N}} \cdot (k_B T / V M_{\mathrm{sat}})$ so that
$\mathcal{N} J \gg V B M_{\mathrm{sat}} \gg k_B T$. Clearly, finite-size effects
do not affect the HMW outcome in 1D, because $\sqrt{\mathcal{N}}$~$\sim$~$10^4$
is still a very large number for macroscopic $\mathcal{N}$~$\sim$~$N_A^{1/3}$.
In 2D, however, finite size affects the HMW outcome qualitatively, because $\ln
N_A^{2/3}$~$\approx$~$37$, effectively replacing the IR divergence with an
order-of-magnitude reduction. Given that magnetic critical temperatures are
typically a few hundred Kelvins in transition-metal compounds, in $2D$ the HMW
argument only predicts a marked suppression of magnetic order, far from
vanishing.

\subsection{Disorder in the XXZ model}

When any perturbation $\var{H}$ is added to the XXZ
Hamiltonian~(\ref{eq:XXZ-Haml}), the original denominator in
Eq.~(\ref{4:Malmost}) becomes
\begin{equation}
\frac{1}{\mathcal{N}} \sum_{\vb{k} \in \mathrm{BZ}} \frac{1}{(n J \vb{k}^2 / 
k_{\mathrm{BZ}}^2) + \Delta_B(\vb{k}) + \big|B \ev{M_x}_{B,
\var{H}}\big|}\,,
\label{disorder}
\end{equation}
where $\Delta_B(\vb{k}) = V^{-1} \big| \langle [[Q_{\vb{k}}, \var{H}],
Q^{\dag}_{\vb{k}}] \rangle_{B, \var{H}} \big|$ and the index $\var{H}$ indicates
thermal averaging in the presence of $\var{H}$. Thus, formally, any perturbation
that breaks the continuous symmetry of the magnetization, generated by
$Q_{\vb{0}}$, gives a finite $\Delta_B(\vb{0}) \neq 0$, which cuts off the
$\vb{k}^2$ term, invalidating the HMW theorem.

Physically, let us consider the site-disorder model
\begin{equation}
\var{H} = - \sum_i \vb{h}_i \vdot \vb{S}_i\,,
\end{equation}
where $\vb{h}_i$ is a random local field. For definiteness, its planar component
is considered to lie on the unit circle, $|\vb{h}_{i\parallel}|=1$, placing the
model in the RFIO limit~\cite{RFIO5}. The crossover to the Imry-Ma
limit~\cite{Imry75} by widening the amplitude distribution is beyond the scope
of this article. In the present case, $\Delta_B = V^{-1} \big|
\ev{\var{H}_{\parallel}}_{B, \delta} \big|$, where $\var{H}_{\parallel}$
includes planar components of the $\vb{h}_i$ only, and $\delta$ indicates
disorder-averaging. The salient observation is that the disorder effectively
competes with the \emph{interaction} scale $J$ in cutting off the IR divergence:
\begin{equation}
\abs{\qev{M_x}}^2 \leq \frac{\beta J}{\displaystyle \ln(1 + n J / \Delta_0)}
\cdot M_{\mathrm{sat}}^2
\label{4:nejed_temp}
\end{equation}
in 2D, and
\begin{equation}
\abs{\qev{M_x}}^2 \leq \frac{2}{\pi} \sqrt{\frac{\Delta_0}{n J}} \cdot \beta J
\cdot M_{\mathrm{sat}}^2\,
\end{equation}
in 1D. The double angular brackets denote the Bogoliubov quasi-average $\qev{M}
= \lim_{B \to 0} \lim_{V \to \infty} \ev{M}_B$. A similar pattern is observed as
with finite-size effects, with the role of the size of the system taken over by
the ratio between the magnetic coupling and the disorder scale. This ratio is
evidently much smaller than Avogadro's number, so it is easy to imagine a 2D
system with a small amount of disorder, say $n J /
\Delta_0$~$\sim$~$100$, for which IR fluctuations suppress $T_c$ by less than
an order of magnitude below $J$.

The competition between disorder and finite-size effects appears in
Eq.~(\ref{disorder}), before the integration which introduces dimensionality.
Disorder cuts off the quadratic term, meaning that dirty samples can be larger
and still avoid the asymptotic HMW regime, irrespective of dimension. In the
context of simulations, this result is a model-independent validation of the
observation that weak disorder, or even numerical error, efficiently stabilizes
the order parameter in 2D.

In a layered system, the order parameter can also be restored by a weak
interlayer coupling $J_{\perp}\ll J$, for which the denominator in
Eq.~(\ref{4:nejed_temp}) similarly reads $\ln(1 + J / J_{\perp})$. Suppression
of the isotropic 3D $T_c$ by weak interlayer coupling competes with the 2D
disorder effect on an equal footing, because the ratio $J / J_{\perp}$ can
easily be both greater and smaller than $n J / \Delta_0$.

Because the competition between mechanisms occurs in the denominator of the
sum~(\ref{disorder}), which is itself in the denominator of
Eq.~(\ref{4:Malmost}), the \emph{largest} scale ends in the numerator. Suppose,
for illustration, that two mechanisms are at work in the same sample, one of
which would limit $kT_c<\Delta_1$, and the other $kT_c<\Delta_2$, if acting
alone. The two acting together have the net effect
\begin{equation}
kT_c<\max(\Delta_1,\Delta_2)\,,
\end{equation}
because the larger scale gaps the other. This result runs contrary to the naive
impression that the ``stricter'' criterion should be applied.

\section{Superconductivity}

\subsection{Interaction in real space}

We consider a layered two-dimensional system with localized impurities,
electron-electron, and electron-phonon interactions, whose Hamiltonian is
\begin{equation}
\Haml = H_0 + V_{\mathrm{int}}\,.
\label{SC-Haml}
\end{equation}
The non-interacting part is
\begin{equation}
H_0 = \sum_{\vb{k} \sigma} \epsilon_{\vb{k} \sigma} c^{\dag}_{\vb{k} \sigma}
c_{\vb{k} \sigma} + \sum_{\vb{\varrho} \sigma_1 \sigma_2}
\varepsilon_{\vb{\varrho} \sigma_1 \sigma_2} f^{\dag}_{\vb{\varrho}
\sigma_1}
f_{\vb{\varrho} \sigma_2} + \sum_{\vb{k} \lambda} \hbar \omega_{\vb{k}
\lambda} b^{\dag}_{\vb{k} \lambda} b_{\vb{k} \lambda}\,,
\end{equation}
where the three terms refer respectively to mobile carriers, impurities at fixed
arbitrary positions $\vb{\varrho}$, and phonons. $V_{\mathrm{int}}$ contains all
possible hybridizations and interactions among them, subject to the limitation
that they are local in real space and time, i.e.\ admit the usual fermion
continuity equation. From the microscopic point of view, this limitation is
quite mild, because all bare interactions considered in standard many-body
theory are local in this sense.

\subsubsection{The HMW argument.} {The operators to be used in
Bogoliubov's inequality~(\ref{Bglj-ineq}) are:}
\begin{eqnarray}
Q_{\vb{k}} &= \sum_{\vb{R} \sigma} \Elr^{- \iu \vb{k} \vdot \vb{R}}
c^{\dag}_{\vb{R} \sigma} c_{\vb{R} \sigma} + \sum_{\vb{\varrho} \sigma}
\Elr^{- \iu \vb{k} \vdot \vb{\varrho}} f^{\dag}_{\vb{\varrho} \sigma}
f_{\vb{\varrho} \sigma}\,, \label{Qk-probe}\\ A_{\vb{k},\vb{r}} &=
\mathcal{N}^{-1} \sum_{\vb{R}} \Elr^{\iu \vb{k} \vdot \vb{R}} c_{\vb{R}
\sigma_1} c_{\vb{R} + \vb{r} \sigma_2}\,, \label{SC-better-probe}
\end{eqnarray}
where $\mathcal{N}$ is the number of unit cells within the 2D layers, while
$\sigma_1$, $\sigma_2$, and $\vb{r}$ are fixed and arbitrary. Clearly,
$Q_{\vb{k}}$ is a modulation of the number operator, while the commutator of
$A_{\vb{k}}$ with $Q_{\vb{k}}$ produces the microscopic gap operator
\begin{equation}
[A_{\vb{k},\vb{r}}, Q_{\vb{k}}] = (1 + \Elr^{- \iu \vb{k} \vdot \vb{r}})
\Delta^{\vb{K}}_{\sigma_1 \sigma_2}(\vb{r})\,.
\end{equation}
The operator $\Delta^{\vb{K}}_{\sigma_1 \sigma_2}(\vb{r})$ can be summed over
$\vb{K}$, $\vb{r}$, $\sigma_1$, and $\sigma_2$ to obtain every possible
(singlet, triplet, $s$, $p$, $d_{x^2-y^2}$, pair momentum $\vb{K}$, etc.)
superconducting gap operator $\Delta$, thus generalizing previous
results~\cite{Gelfert,supra1,supra2,supra3,supra4,supra5}. Summing Bogoliubov's
inequality~(\ref{Bglj-ineq}) over $\vb{k}$, one obtains the SC analogue of
Eq.~(\ref{4:Malmost}):
\begin{equation}
\abs{\ev{\Delta^{\vb{K}}_{\sigma_1 \sigma_2}(\vb{r})}_{\kappa}}^2 \leq
\frac{\ell \, \beta}{\displaystyle \frac{2}{\mathcal{N}} \sum_{\vb{k}}
\frac{\abs{1 + \Elr^{\iu \vb{k} \vdot \vb{r}}}^2}{(\epsilon_{\mathrm{IR}} k^2
/  k_{\mathrm{BZ}}^2) + \abs{\kappa} \mathcal{D}_{\kappa}}}\,,
\label{6:tilde-delta-almost}
\end{equation}
where $\ell$ is the number of layers, $\kappa$ is the symmetry-breaking field
conjugate to $\Delta$, same as $B$ in Eq.~(\ref{4:Malmost}), and
$\mathcal{D}_{\kappa}$ is a corresponding expectation value of $\Delta$. The
infrared energy scale $\epsilon_{\mathrm{IR}}$ is determined by the bound
$(\mathcal{N} \ell)^{-1} \big| \langle [[Q_{\vb{k}}, \Haml], Q^\dag_{\vb{k}}]
\rangle \big| \leq \epsilon_{\mathrm{IR}} k^2 /  k_{\mathrm{BZ}}^2$. It is a
sum of two positive terms: one due to particle dispersions and another from
hybridization with impurities. The particle-dispersion term is always finite,
while the impurity hybridization $V_{\sigma_1\sigma_2}^{\vb{\varrho}\vb{R}}
(f^{\dag}_{\vb{\varrho} \sigma_1}c_{\vb{R} \sigma_2} + \mathrm{c.c.})$ must fall
off sufficiently rapidly, $\mathcal{N}^{-1} \sum_{\vb{\varrho},\vb{R}}
\abs{\vb{\varrho} - \vb{R}}^2 \abs{V_{\sigma_1 \sigma_2}^{\vb{\varrho}
\vb{R}}} < \infty$, for the impurity term to be finite. The electron-electron
and electron-phonon interactions drop out in $[Q, \Haml]$ because they are local
in space. Because $\abs{1 + \Elr^{\iu \vb{k} \vdot \vb{r}}}^2\neq 0$ at
$\vb{k}=0$, the IR divergence is unchecked in 1D and 2D when $\kappa\to 0$, so
the HMW theorem follows.

\subsubsection{Physical considerations.}

Technically, hopping terms and couplings to them are non-local, but even such 
interactions do not invalidate the above conclusion. They are merely subject to
the same caution as impurity hybridization, namely that their range falls off
fast enough. Hence the HMW theorem in its original formulation is quite robust
with respect to model variations. To ascertain how physically relevant it is, we
rewrite the bound~(\ref{6:tilde-delta-almost}) in a more intuitive form:
\begin{equation}
\abs{\ev{\Delta^{\vb{K}}_{\sigma_1 \sigma_2}(\vb{r})}_{\kappa}} \leq
\sqrt{\frac{T_{\mathrm{HMW}}}{T}} \cdot 1\,,
\end{equation}
where $T_{\mathrm{HMW}}$ is the HMW temperature and $1$ is the saturation value
of $\ev{\Delta^{\vb{K}}_{\sigma_1 \sigma_2}(\vb{r})}$. Specializing to 2D,
assuming a parabolic dispersion, and taking some care with numerical factors,
one finds $k_B T_{\mathrm{HMW}} \approx (\ell / 4) (\hbar^2 k_{\mathrm{BZ}}^2 /
2 m_*) \ev{n} / \ln(k_{\mathrm{BZ}} L)$, where $\ev{n} =
\ev{N} / (\mathcal{N} \ell)$ is the total number of carriers per unit cell and
$L$ is the linear size of the sample. Therefore, the HMW bound is physically
relevant only for temperatures $T > T_{\mathrm{HMW}}$, when enough IR modes are
excited to suppress the SC order. The conclusion, which applies to all previous
HMW arguments concerning
SC~\cite{Hohenberg,Sadovnikov,Gelfert,supra1,supra2,supra3,supra4,supra5} as
well, is that in 2D
\begin{equation}
T_c < 4 T_{\mathrm{HMW}} = \ell \cdot T_F / \ln(k_{\mathrm{BZ}} L)\,,
\end{equation}
where $T_F$ is the Fermi temperature. Inserting $T_F\sim 10^4$~K,
$k_{\mathrm{BZ}}\sim$~\AA$^{-1}$, and $L\sim 1$~cm, one finds the bound to be in
the $\sim 500$~K range, which is comparable to the bound $k_B T_c < E_F / 8$,
obtained by considering phase stiffness in an infinite
system~\cite{alt-Tc-bounds}.

The true significance of the finite-size effect becomes evident in the converse
exercise: $k_{\mathrm{BZ}}L<\exp(T_F/T_c)$. One would need
$L$~$\sim$~$10^{33}$~m, much larger than the observable universe
($\sim$~$10^{27}$~m), for $T_c$ in a single layer to be forced below
$\sim$~$100$~K, the value observed in optimally doped cuprates.

\subsection{Interaction in momentum space} \label{momentum-section}

The critical junction in the above derivation is that the commutator
$[Q_{\vb{k}}, V_{\mathrm{int}}]$ vanishes, because interactions are local and
therefore commute with the local number operators appearing in
$Q_{\vb{k}}$~(\ref{Qk-probe}). The f-sum rule $(\mathcal{N} \ell)^{-1}
\big\langle [[Q_{\vb{k}}, \Haml], Q_{\vb{k}}^{\dag}] \big\rangle \lesssim
\epsilon_{\mathrm{IR}} k^2 /  k_{\mathrm{BZ}}^2$ follows, which is physically
the statement that boosting the electrons by $\hbar \vb{k}$ results in a
second-order-in-$\vb{k}$ energy increase of the form $\ell\mathcal{N}
\epsilon_{\mathrm{IR}} k^2 /  k_{\mathrm{BZ}}^2$. As a quick way to derive
this quadratic dependence, write $\hat{Q}_{\vb{k}} = \hat{N} - \iu \vb{k}
\vdot \hat{\vb{X}} + \mathcal{O}(k^2)$, where $\hat{N} = \hat{Q}_{\vb{0}} =
\sum_{\vb{R}} \hat{n}_{\vb{R}}$ and $\hat{\vb{X}} = \sum_{\vb{R}} \vb{R} \,
\hat{n}_{\vb{R}}$. The double commutator is immediately
\begin{equation}
{}[[Q_{\vb{k}}, \Haml], Q_{\vb{k}}^{\dag}] = [[\vb{k} \vdot\hat{\vb{X}},
\Haml], \vb{k} \vdot\hat{\vb{X}}] + \mathcal{O}(k^3)\,,
\end{equation}
because $[\hat{N}, \Haml] = 0$ of course follows from local number-conservation.
Therefore, the HMW argument is model-independent, as long as the interactions,
or disorder, conserve the particle number locally, and boosting the electrons of
the system increases the energy by an extensive amount. All later proofs thus
necessarily reproduce its physically most counter-intuitive aspect, that one
probes high-energy IR fluctuations of the non-interacting Fermi sea, instead of
the superconducting mechanism.

Having understood that, we develop an alternative argument which probes
fluctuations associated with Cooper pairing, as expressed in the reduced BCS
Hamiltonian:
\begin{equation}
\Haml = \sum_{\vb{k} \sigma} \epsilon_{\vb{k}} c^{\dag}_{\vb{k} \sigma}
c_{\vb{k} \sigma} + \frac{1}{\mathcal{N}} \sum_{\vb{k}_1 \vb{k}_2} V_{\vb{k}_1
\vb{k}_2} c^{\dag}_{\vb{k}_1 \uparrow} c^{\dag}_{- \vb{k}_1 \downarrow} c_{-
\vb{k}_2 \downarrow} c_{\vb{k}_2 \uparrow}\,.
\label{bcsham}
\end{equation}
The reduced BCS Hamiltonian breaks the standard HMW argument because the
interaction is non-local in real space, violating local continuity. It causes
$\epsilon_{\mathrm{IR}}$ in Eq.~(\ref{6:tilde-delta-almost}) to diverge with
$(k_{\mathrm{BZ}} L)^2$ in the thermodynamic limit, as typical for long-range
interactions. In the remainder of this section, we adapt the technical steps of
the original argument to the BCS pair interaction, leaving the physical
discussion to the end.

The operator $Q$ is constructed to commute with the kinetic energy now, and $A$
adjusted so that $[A, Q]$ gives the gap operator as before:
\begin{eqnarray}
Q_{\vb{R}} = \sum_{\vb{k}} \Elr^{\iu \vb{k} \vdot \vb{R}}
\mleft(c^{\dag}_{\vb{k} \uparrow} c^{\vphantom{\dag}}_{\vb{k} \uparrow} +
c^{\dag}_{- \vb{k} \downarrow} c^{\vphantom{\dag}}_{- \vb{k}
\downarrow}\mright)\,, \\ A_{\vb{R},\vb{K}} = \mathcal{N}^{-1} \sum_{\vb{k}}
\Elr^{- \iu \vb{k} \vdot \vb{R}} S_{\vb{K}}(\vb{k}) c_{\vb{k} + \vb{K}
\uparrow} c_{- \vb{k} \downarrow}\,, \\ {}[A_{\vb{R},\vb{K}}, Q_{\vb{R}}] = (1
+ \Elr^{\iu \vb{K} \vdot \vb{R}}) \Delta_{\vb{K}}\,,
\end{eqnarray}
where $S_{\vb{K}}(\vb{k})$ is a Cooper-pair structure factor which allows a
finite pair momentum $\vb{K}$ for the sake of generality.

The analogue of Eqs.~(\ref{4:Malmost})~and~(\ref{6:tilde-delta-almost}) reads,
in any dimension,
\begin{equation}
\abs{\ev{\Delta_{\vb{K}}}_{\kappa}}^2 \leq I_{\vb{K}} \cdot \beta
\mleft\{\frac{1}{\mathcal{N}} \sum_{\vb{R}} \frac{\abs{1 + \Elr^{\iu \vb{K}
\vdot \vb{R}}}^2}{\mathcal{V}_{\kappa}(\vb{R}) + \abs{\kappa}
\mathcal{D}_\kappa}\mright\}^{-1}\,,
\label{delta-mom-space}
\end{equation}
where $I_{\vb{K}} = \mathcal{N}^{-1} \sum_{\vb{k}} \abs{S_{\vb{K}}(\vb{k})}^2$
is a kinematic factor. In contrast to small values of $\vb{k}$ in the HMW
argument, there is nothing special about small $\vb{R}$ here, so the total sum
does not diverge in the thermodynamic limit in any dimension. Thus,
Eq.~(\ref{delta-mom-space}) does not constrain the order parameter to vanish.
Nevertheless, the upper bound for $T_c$ it provides is sharper than in the
original formulation for samples of reasonable size when the Cooper pairs are
small, as we show below.

The term $\mathcal{V}_{\kappa}(\vb{R})$ is an average of the interaction
operator,
\begin{equation}
\mathcal{V}_{\kappa}(\vb{R}) = \frac{8}{\mathcal{N}^2} \sum_{\vb{k}_1
\vb{k}_2} \sin^2\mleft[(\vb{k}_1 - \vb{k}_2) \vdot \vb{R} / 2\mright] \,
\abs{V_{\vb{k}_1 \vb{k}_2}} \langle c^{\dag}_{\vb{k}_1 \uparrow} c^{\dag}_{-
\vb{k}_1 \downarrow} c_{- \vb{k}_2 \downarrow} c_{\vb{k}_2 \uparrow}
\rangle_{\kappa}\,,
\end{equation}
while $\mathcal{D}_\kappa$ averages the order parameter like in
Eq.~(\ref{6:tilde-delta-almost}). A more transparent form is obtained after
angularly averaging $\abs{1 + \Elr^{\iu \vb{K} \vdot \vb{R}}}^2$ in 2D and 3D,
\begin{equation}
\abs{\qev{{\Delta}_{\vb{K}}}} \leq \sqrt{\frac{T_{\mathrm{pair}}}{T}} \cdot
\Delta^{\vb{K}}_{\mathrm{sat}}\,, \label{B:newineq}
\end{equation}
where $\Delta^{\vb{K}}_{\mathrm{sat}} = \abs{\mathcal{N}^{-1} \sum_{\vb{k}}
S_{\vb{K}}(\vb{k}) \ev{c_{\vb{k} + \vb{K} \uparrow} c_{- \vb{k}
\downarrow}}_{T=0}}$ is the saturation value of the order parameter at $T =
0$, while $T_{\mathrm{pair}}$ is the upper bound temperature
\begin{equation}
k_B T_{\mathrm{pair}} = \frac{I_{\vb{K}}}{\displaystyle
(\Delta^{\vb{K}}_{\mathrm{sat}})^2} \cdot \frac{8}{\mathcal{N}^2}
\sum_{\vb{k}_1 \vb{k}_2} \abs{V_{\vb{k}_1 \vb{k}_2}} \langle
c^{\dag}_{\vb{k}_1 \uparrow} c^{\dag}_{- \vb{k}_1 \downarrow} c_{- \vb{k}_2
\downarrow} c_{\vb{k}_2 \uparrow} \rangle\,. \label{B:ximax}
\end{equation}
In the thermodynamic limit, both $\mathcal{V}_{\kappa}(\vb{R})$ and $k_B
T_{\mathrm{pair}}$ tend to \emph{intensive} constants. Thus the upper
bound~(\ref{B:newineq}) is independent of the size of the sample. For more
general interactions, the above is easily generalized to
\begin{equation}
k_B T_{\mathrm{pair}} = \frac{I_{\vb{K}}}{\displaystyle
(\Delta^{\vb{K}}_{\mathrm{sat}})^2} \cdot \frac{8}{\mathcal{N}^2} \sum_{\vb{q}
\vb{k}_1 \vb{k}_2} \sum_{\sigma_1 \sigma_2} \abs{V_{\vb{q}}} \langle
c^{\dag}_{\vb{k}_1 + \vb{q} \sigma_1} c^{\dag}_{\vb{k}_2 - \vb{q} \sigma_2}
c_{\vb{k}_2 \sigma_2} c_{\vb{k}_1 \sigma_1} \rangle\,.
\end{equation}

\subsubsection{Evaluation of the upper bound temperature.} First, we estimate
the ratio
\begin{equation}
\frac{I_{\vb{K}}}{\displaystyle (\Delta^{\vb{K}}_{\mathrm{sat}})^2} =
\frac{\displaystyle \mathcal{N}^{-1} \sum\nolimits_{\vb{k}}
\abs{S_{\vb{K}}(\vb{k})}^2}{\displaystyle \Big| \mathcal{N}^{-1}
\sum\nolimits_{\vb{k}} S_{\vb{K}}(\vb{k}) \ev{c_{\vb{k} + \vb{K} \uparrow}
c_{- \vb{k} \downarrow}}_{T=0} \Big|^2}\,.
\end{equation}
It is reasonable to set the Cooper-scattering structure factor $S_{\vb{K}}$ to
zero outside a shell of thickness $\var{k}_{\mathrm{gap}}$ on the surface of the
Fermi sphere, leading to
\begin{equation}
\frac{I^{\vb{K}}}{\displaystyle (\Delta^{\vb{K}}_{\mathrm{sat}})^2} \sim
\frac{k_{\mathrm{BZ}}}{\var{k}_{\mathrm{gap}}} \sim \xi_P\,,
\end{equation}
where $\xi_P$ is the size of a Cooper pair (Pippard scale) in units of the
lattice constant. Here, we have taken into account that $\ev{c_{\vb{k} +
\vb{K} \uparrow} c_{- \vb{k} \downarrow}}_{T=0}\propto \Delta_{SC} /
\sqrt{(\epsilon_{\vb{k}} - E_F)^2 + \Delta_{SC}^2}$~$\sim$~$1$ within
$\var{k}_{\mathrm{gap}}$ from the Fermi surface, where the microscopic gap
$\Delta_{SC}\neq 0$.

Next, we take the BCS interaction to be the usual schematic $-V_0$ within a
range of $\pm\hbar\omega_D$ around $E_F$, and introduce the dimensionless
coupling constant $\lambda = A g(E_F) V_0$, where $A$ is the surface of the unit
cell and $g(E_F)$ the (intensive) level density at the Fermi energy. Then
\begin{equation}
k_B T_{\mathrm{pair}} \sim \xi_P \cdot 8 \cdot V_0 \cdot [A g(E_F) \hbar
\omega_D]^2 \sim 10 \xi_P \cdot \frac{\lambda^2 (\hbar \omega_D)^2}{V_0}\,.
\label{kbTstar}
\end{equation}

When the actual numbers for some classical 3D superconductors~\cite{aluminium}
are inserted, one finds that the intensive bound~(\ref{kbTstar}) still allows 2D
SC by a wide margin, primarily because Cooper pairs are so large,
$\xi_P$~$\sim$~$1000$. For hafnium, $A g(E_F) \approx 0.8$~eV$^{-1}$, $\lambda
\approx 0.14$ and $\hbar \omega_D \approx 22$~meV, which gives
$T_{\mathrm{pair}} \approx 4000$~K, while for aluminium, $A g(E_F) \approx
0.5$~eV$^{-1}$, $\lambda \approx 0.4$ and $\hbar \omega_D \approx 36$~meV,
giving $T_{\mathrm{pair}} \approx 15000$~K. The measured values of the SC $T_c$
are $0.13$~K and $1.2$~K, respectively.

A similar conclusion pertains to high-$T_c$ superconducting cuprates, even
though their Cooper pairs are small, $\xi_P$~$\sim 1$--$3$, and the critical
temperatures are much larger, in the $\sim$~$100$~K range. The SC mechanism for
them is not known at present. There is broad consensus that it is not phononic,
but the schematic BCS Hamiltonian is still applicable if the SC involves Cooper
pairs. One can replace $\hbar \omega_D$ with an electronic scattering scale
$\sim$~$100$~meV, $V_0$ with an effective Mott or charge-transfer local scale
$U_{\mathrm{eff}}\sim 3$--$5$~eV, and $\lambda$ with $\sim$~$1$ for a strong
coupling. One obtains an upper bound $T_{\mathrm{pair}}$~$\sim$~$1000$~K, which
is surprisingly reasonable, given the crudity of the estimates.

\section{Discussion}

The textbook Cooper pair attraction is not local in real space, so it is not a
true microscopic interaction. Because it retains only the one scattering channel
of the full interaction which gives rise to the SC instability, it is the
effective macroscopic interaction, which dominates all other scattering channels
when the transition is approached from above. Hohenberg's original argument,
which relied only on the equation of continuity, did not make any assumption
about the microscopic origin of the SC order parameter.

The explicit invocation of Cooper pairing in this work opens the way for the HMW
argument to probe the SC mechanism. It chooses a specific macroscopic
(low-temperature) realization of the generic microscopic (high-temperature)
original argument. Unlike the original argument, this specialization does not
include e.g.\ polaronic models, where the SC order parameter appears by
bound-pair formation, instead of Cooper scattering.

The Cooper-pair attraction used here is schematic. It should not be surprising
that it gives a rather loose bound for $T_c$, because it has to accommodate all
possible SC $T_c$'s based on the Cooper instability. To make the same point
conversely, if such a purely schematic interaction had given a limit on $T_c$
much lower than the observed values, that would have indicated an error in the
reasoning.

The same approach can in principle be improved. Specifics of the crystal
symmetry and the scattering channel can be introduced via the Cooper-pair form
factor. The modulation $Q_{\vb{R}}$ which probes the fluctuations of the
scatterer density can also be adapted to reflect a particular idea or
measurement. One can envisage the HMW argument gradually migrating, with
successive refinements, from the mathematical realm of no-go theorems to the
physical reality of concrete materials and samples. {In this
context, our estimate being closer to observation for cuprates than for metals
could mean that neglecting retardation effects in the interaction~(\ref{bcsham})
is more realistic for the former, as argued in particular by
Anderson~\cite{Anderson07}.}

The suppression of 3D SC in thin films of elemental BCS superconductors proceeds
by a different mechanism than in the HMW argument. The minimal film thickness
for Meissner shielding in clean superconducting metals is an effective
penetration depth $\lambda_{\mathrm{eff}} =
\sqrt[3]{\lambda_L^2\xi_P}$, where $\lambda_L$ is the London penetration
depth~\cite{Pippard53}. In aluminium, $\lambda_{\mathrm{eff}}\sim 1000$~\AA, or
$250$ lattice spacings, indicating that the sample needs to be macroscopic in
all three dimensions in order to provide enough phase space for SC. By contrast,
a SC cuprate layer of single-unit-cell thickness has a perfect Meissner
response~\cite{cuprate-monolayer1}.

{It is well known that the Berezinskii-Kosterlitz-Thouless (BKT)
mechanism~\cite{Berezinskii71,Berezinskii72,Kosterlitz72,Kosterlitz73} allows
for sharp phase transitions in 2D with power-law decays in the spatial
correlation functions, instead of their saturating asymptotically at finite
values as in true long-range order. BKT fluctuations associated with vortex
pairs appear in both low- (BCS) and high-T$_c$ (cuprate) SC thin films, with the
BKT transition always observed as a separate phenomenon within the SC
regime~\cite{Mondal12,Artemenko92}. Hence the BKT mechanism is not sufficient to
stabilize 2D SC in general, although it may affect properties of the SC state
near the critical temperature~\cite{Mondal12}.}

\section{Conclusion}

The main result of the present work is that no special pleading, such as
substrate effects or the Berezinskii-Kosterlitz-Thouless mechanism, is necessary
in 2D to account for either magnetism or superconductivity in light of the HMW
theorem. The precise meaning of this statement is different in the two cases.

For magnetism, the infrared fluctuations affect $T_c$ in 2D samples of
reasonable size sufficiently to reduce it by an order of magnitude. In the
presence of disorder or weak interlayer coupling, however, these fluctuations
become significantly less effective in suppressing magnetic order. Infrared
fluctuations are thus not a universal explanation for the reduction of $T_c$ in
any given 2D system. It is not possible to state a priori why $T_c>0$ in a given
2D film, nor how much it should be reduced relative to its value in the bulk. A
particular analysis of the sample at hand, comparing finite-size effects with
interlayer-coupling and disorder scales, is required. In general, the mechanism
which suppresses $T_c$ the least will prevail.

For SC, the theorem is ineffective in 2D for all physically possible sample
sizes. The reason for the ineffectiveness is that the cost of infrared
fluctuations in the local number density is set by the kinetic-energy scale,
which is very high in metals. These infrared modes are not sufficiently excited
at SC onset temperatures to affect the SC order parameter in 2D, making the
original HMW bound physically void. Bulk values of $T_c$ observed in atomically
flat films~\cite{cuprate-monolayer1} and exfoliated
layers~\cite{cuprate-monolayer4} should not be surprising at all. Rather, the
surprise is that they were considered puzzling for so long.

The upper limit for the SC $T_c$ we find is in the thousand-Kelvin range, so its
practical significance is qualitative. While the original HMW theorem in
principle allows a 2D SC order parameter as a finite-size effect, the present
formulation eliminates the worry that infrared fluctuations will compromise
Cooper pairing as a particular mechanism to realize it. No physically reasonable
value for the SC $T_c$ in 2D is precluded by any known version of the HMW
argument as of this writing. We conclude that all physical and chemical
considerations for the appearance of 2D SC take precedence over dimensionality.
We hope that this result will stimulate the search for high-temperature SC in
fabricated 2D materials.

\begin{acknowledgments}
A careful reading of the manuscript by O.~S.~Bari\v{s}i\'c is gratefully
acknowledged. This work was funded by the Croatian Science Foundation under
Project No. IP-2018-01-7828.
\end{acknowledgments}

\bibliographystyle{unsrt}
\bibliography{hmw_v7a.bib}

\end{document}